%% file: main.tex
\documentclass[conference]{IEEEtran}
\IEEEoverridecommandlockouts
\usepackage{cite}
\usepackage{amsmath,amssymb,amsfonts}
\usepackage{algorithmic}
\usepackage{graphicx}
\usepackage{textcomp}
\usepackage{xcolor}
\def\BibTeX{{\rm B\kern-.05em{\sc i\kern-.025em b}\kern-.08em
    T\kern-.1667em\lower.7ex\hbox{E}\kern-.125emX}}

\usepackage{algorithmic}
\usepackage{graphicx}
\usepackage{textcomp}
\usepackage{xcolor}
\usepackage{tabularx}
\usepackage{tabularx,booktabs,ragged2e}
\usepackage{subcaption}
\usepackage{listings}
\usepackage{makecell}
\usepackage{url}
\usepackage{listings}
\usepackage{amsmath}
\usepackage{comment}
\usepackage{tikz}

\usepackage{hyperref}
\usepackage{xcolor}

\begin{document}

\title{NDNSD: Service Publishing and Discovery in NDN}

\author{\IEEEauthorblockN{Saurab Dulal}
\IEEEauthorblockA{University of Memphis\\ sdulal@memphis.edu}\vspace{-0.7cm}
\and
\IEEEauthorblockN{Lan Wang}
\IEEEauthorblockA{University of Memphis\\ lanwang@memphis.edu}\vspace{-0.7cm}
}





\maketitle

\pagestyle{plain}

\input{abstract}

\begin{IEEEkeywords}
Information Centric Networking, Named Data Networking, Service Discovery, NDN Synchronization
\end{IEEEkeywords}

\input{introduction.tex}
\input{background.tex}
\input{related-works.tex}
\input{system-design.tex}

\input{evaluation.tex}
\input{conclusion.tex}

\bibliographystyle{IEEEtran}
\bibliography{refs}

\end{document}

%% file: abstract.tex
\begin{abstract}

Service discovery is a crucial component in today's massively distributed applications.   
 In this paper, we propose NDNSD -- a fully distributed and general-purpose service discovery protocol for Named Data Networking (NDN). By leveraging NDN's data synchronization capability, NDNSD offers a high-level API for service publishing and discovery. We present NDNSD's main design features including hierarchical naming, service information specification, and service accessibility. 
We also implemented two other discovery schemes, one reactive and one proactive, and compared them with NDNSD. Our evaluation shows that NDNSD achieves (a) lower latency, lower overhead, and same reliability compared to the reactive scheme, and (b) comparable latency, lower overhead at larger scale, and higher reliability compared to the proactive scheme. 

\end{abstract}

%% file: introduction.tex
\section{Introduction}
\label{introduction}
Modern applications rely on many services, e.g., cloud computing, file storage, and databases, to function properly.  These services are provided by not only stationary servers, but also mobile devices offering computing and data service. 
\emph{\textbf{Service discovery}} is the process of finding desired service(s) in a distributed environment.

TCP/IP has a plethora of protocols
(\cite{unpn_jeronimo2003upnp, bonjour_boucadair2013universal, llsd, jini, slp_guttman1999service})
at different layers to facilitate service discovery.  These protocols have a major limitation stemming from the inability of the network layer to recognize names used by the application layer (\cite{shang2016challenges, shang2016named}). For example, a printer application cannot identify itself as \textit{/edu/abc/library/printer1} to the network, but instead has to rely on its IP address and port number. This creates a semantic mismatch between the two layers and thus requires either a dedicated server (e.g., the resource directory in CoAP) or a name resolution service such as DNS. 
This dependency is cumbersome for decentralized applications or edge applications.  Popular edge applications such as AWS IoT \cite{aws-iot}, Google Chromecast~\cite{chromecast}, and Azure Sphere \cite{azure-sphere} mostly depend on the cloud, not a local device, for service registration and discovery. 
This design incurs extra delay for edge devices to discover services residing in close vicinity. More seriously, a disruption in the connectivity to the cloud will disrupt the whole application. 
  
Named Data Networking (NDN) is a new data-centric Internet architecture~\cite{zhang2014named}.  It uses application-level names directly in the network layer, so there is no need to resolve the application names into addresses and ports (and there are no addresses in the network layer).  Moreover, NDN's distributed dataset synchronization protocols~\cite{moll2021survey}, i.e., \textit{Sync}, can make the discovery process fully distributed by directly synchronizing the service information among the participating nodes, thereby eliminating the requirement of a centralized entity to facilitate the service discovery process.
Several service discovery solutions~(\cite{mosko2014ccnx, ravindran2013information, ndnlite, amadeo2016ndne, mastorakis2020icedge})
have been proposed for NDN (and its predecessor CCN). However, we found various limitations in these solutions (Section~\ref{related-work}) such as high overhead, being limited to a specific environment, or requiring separate servers to store service information.  

In this paper, we propose NDNSD, 
a fully distributed, general-purpose, and scalable service discovery protocol for NDN. NDNSD offers a sync-based high-level API for publishing and discovering services, obtaining measurement information, and controlling access to service information.
We have implemented NDNSD
and evaluated it 
by comparing it to two other discovery schemes, one reactive and one proactive. Our evaluation shows that NDNSD achieves (a) lower latency, lower overhead, and same reliability compared to the reactive scheme, and (b) comparable latency, lower overhead at a larger scale, and higher reliability compared to the proactive scheme. 


%% file: background.tex
\section{Background and Related Work} 
\label{background-related-worlk}

\subsection{NDN}
\vspace{-5pt}
\textit{Named Data Networking (NDN)} is an evolving data-centric Internet architecture. Every piece of content in NDN is named (e.g. /edu/abc/servers/cygnux/info), carried in one or more \textit{data packets}. The content is fetched using \textit{Interests} whose name matches the data name. NDN changes IP's host-centric communication by decoupling data packets from their producers. The decoupling is feasible in NDN because data is signed by its producer at the time of creation and thus its authenticity can be verified by other nodes. Once decoupled, data can be served by any node that stores a copy of it.

An NDN forwarder, e.g., the NDN Forwarding Daemon (NFD) \cite{nfd2022}, implements the network-layer protocols needed for name-based communication. The forwarder consists of three core components: Pending Interest Table (PIT), Content Store (CS), and Forwarding Information Base (FIB). The PIT records incoming Interests, not yet satisfied, and also aggregates them if the same Interest has already been received. The CS caches previously received data packets to satisfy incoming Interests. If an Interest is not satisfied by the CS or matches the PIT, it is forwarded via one or more interfaces with the help of the FIB and a forwarding strategy. Once a data packet is received, it is forwarded to all the incoming interfaces of the matching Interests recorded in the PIT, thus multicast is supported natively in NDN.  In addition, since each data packet follows the reverse path of the matching Interests, routers can measure path performance (e.g., RTT), which enables adaptive forwarding decisions~\cite{yi2013case}. 
\begin{figure}
    \centering
    \includegraphics[scale=0.38, angle=0]{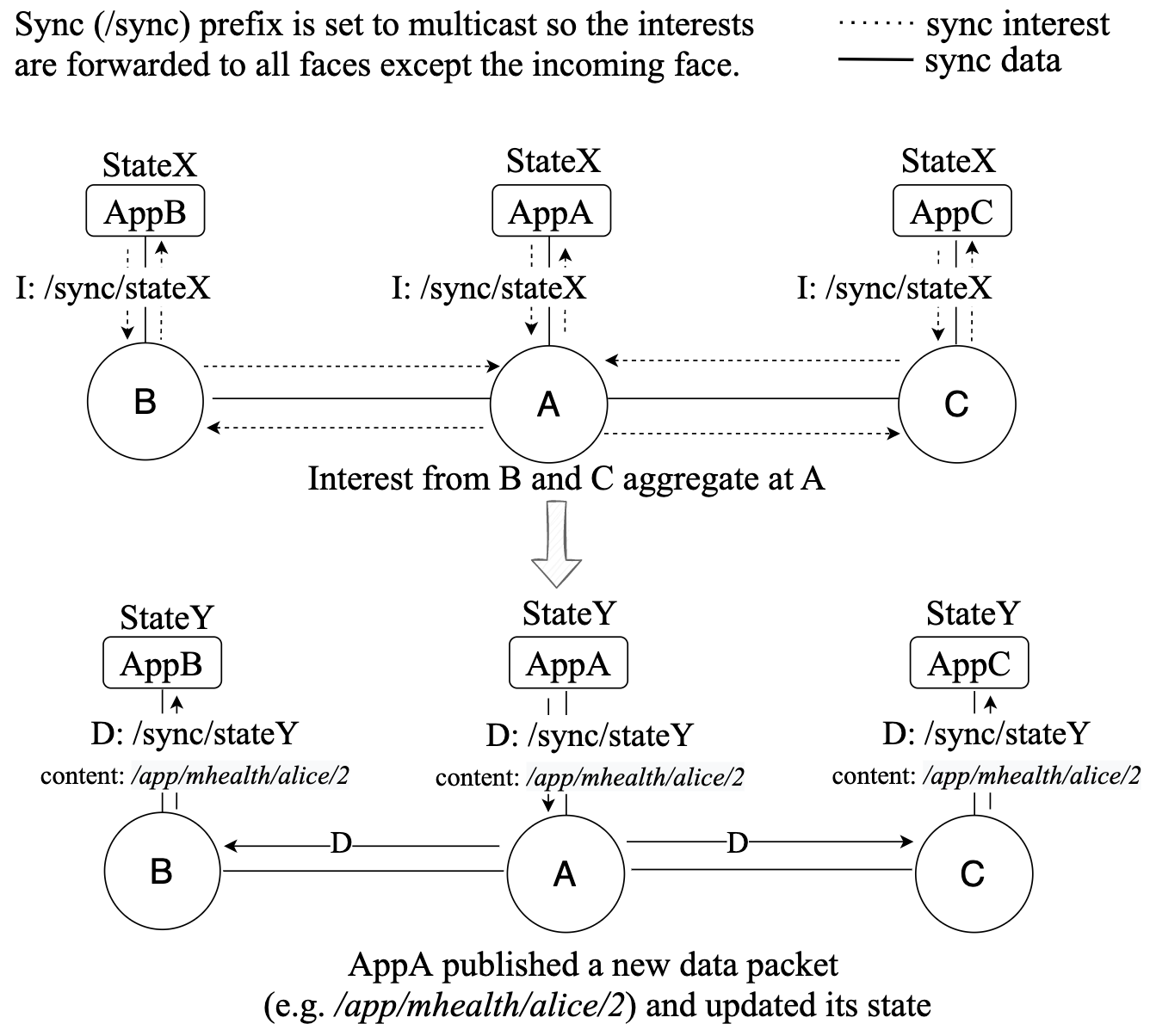}
    \caption{Example of NDN Synchronization Process (``I" and ``D" represent Sync Interest and Sync Data, respectively.)}
    \label{fig:syncronization-process}
    \vspace{-12 pt}
\end{figure}

\subsection{Data Synchronization in NDN (Sync)}
\vspace{-3pt}



NDN synchronization, or \textit{Sync} in short, provides a powerful abstraction above the Interest-Data exchange to facilitate multi-party communication~\cite{moll2021survey}.  Sync ensures that every participating node has up-to-date information of \textit{their shared distributed dataset} by encoding the set of data names in a compact form (i.e., ``state'') and exchanging the state among the nodes using Sync Interests (Figure~\ref{fig:syncronization-process}).
Once a new data packet is published, Sync updates its local state and sends the new state and new data name in a Sync Data packet to all other nodes (Figure~\ref{fig:syncronization-process}), which can now fetch the data using the new data name.

%% file: related-works.tex
\subsection{Related Work}
\label{related-work}
\vspace{-3pt}
Mark Mosko suggested using Sync for service discovery~\cite{mosko2014ccnx}. Devices use a well-defined namespace such as \textit{``/parc/printers/"} to advertise the manifest of service records. However, Mosko's design lacks protocol details, API specifications, and access control. Moreover, it was designed for a different ICN architecture, i.e. CCNx 1.0, as opposed to NDN.
Ravindran et.al. \cite{ravindran2013information} proposed two different service discovery protocols: neighbor discovery protocol (NDP) for locally reachable CCNx nodes neighbors, and Service Publish and Discovery (SPDP) for discovering remote services. SPDP uses a recursive query that propagates hop-by-hop among the reachable adjacencies running SPDP instances. Data containing the service list is aggregated by the respective instances and is sent back to the original requestor. This approach searches for services one hop at a time, so
it may take a long time if services are multiple hops away.

NDNe~\cite{amadeo2016ndne} uses an expanded ring search technique along with the broadcast for service discovery in edge environments.  
The request is first broadcasted to a 1-hop neighbor (TTL is used for hop count).
If no reply is received within the pre-defined timeout, the request is sent to 2-hop neighbors.  This process repeats until either the service is found, or the consumer gives up. 
The expanded ring search can be expensive and may not scale if there is a large number of consumers.

Similar to our work, Mastorakis et. al. 
use Sync among multiple edge computing servers (ECS) to facilitate service discovery~\cite{mastorakis2020icedge}. ECS are special nodes in the network that maintain service information available from service providers. A discovery Interest (e.g. \textit{/discovery/ecs/}) from an application is matched with a suitable service by the closest ECS. However, maintaining ECS is an extra infrastructure requirement for service discovery -- this may be feasible for a particular edge computing application, but not in the general case. 

%% file: system-design.tex
\section{Design}
\label{design}

\begin{figure}
    \centering
    \includegraphics[scale=0.28, angle=0]{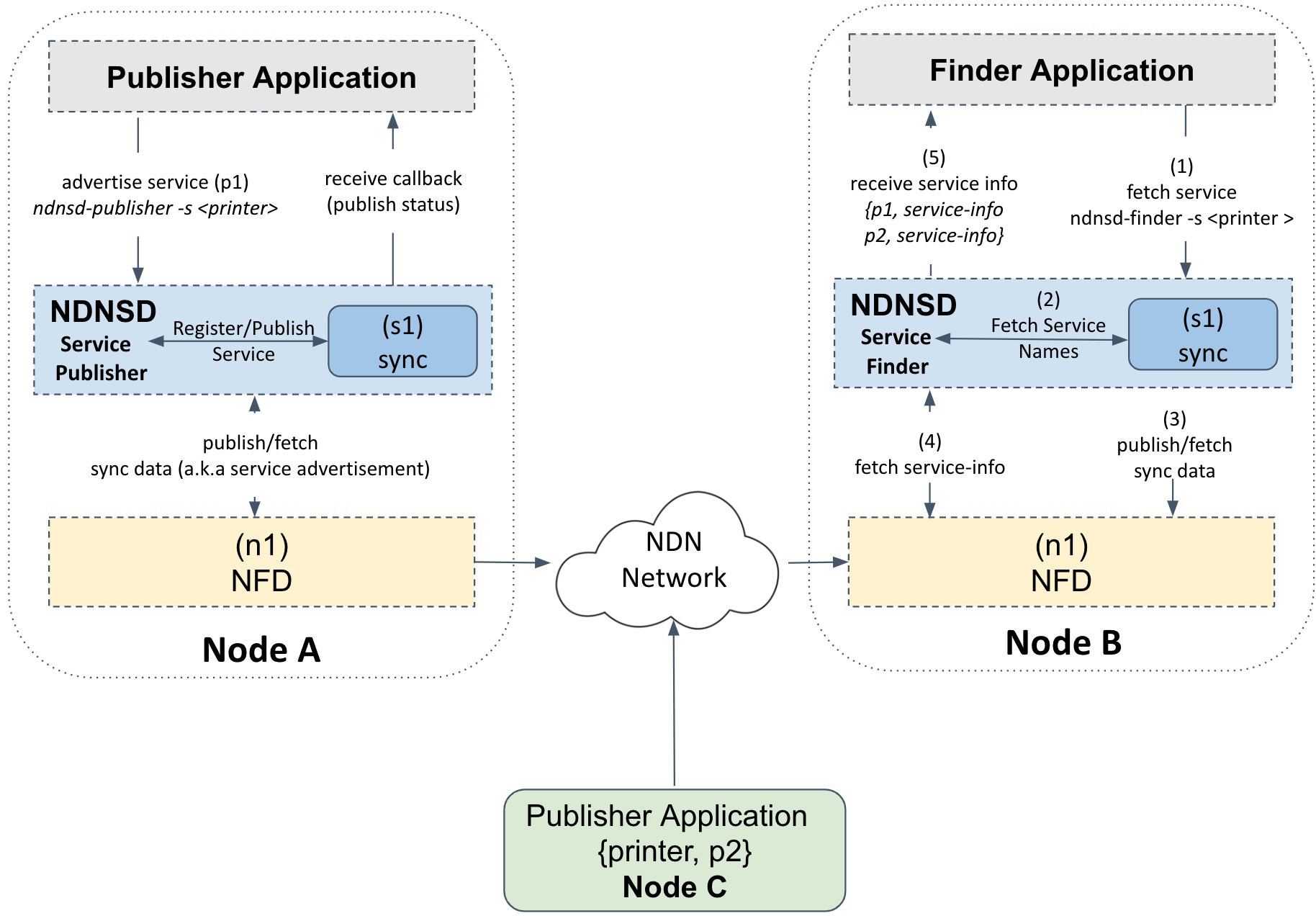}
    \caption[Application workflow showing all layers involved in process]{Application workflow showing all layers involved in service publishing and discovery process}
    \label{fig:workflow}
\end{figure}

We designed NDNSD to be a fully distributed and general-purpose service discovery protocol for NDN that works in a wide range of environments, e.g., LAN, WAN, and IoT. It provides an API for applications to advertise and discover services. Internally, it uses NDN Sync~\cite{moll2021survey} for service announcement and discovery.

We view service discovery as 
a pub-sub problem, i.e. publishing service information and subscribing to such information. 
Participants in this pub-sub system can be one of three types: i) those advertising services via publishing service information, e.g., an NDN repository providing persistent storage service to other devices, ii) those discovering services via subscribing to service information, e.g., a data collection application on a sensor that needs to use a remote persistent storage service, and iii) those doing both, e.g., a group of computers that dynamically share their spare computation resources with each other.  Since Sync provides transport service to NDN applications, it can be used to realize a pub-sub system, as shown by Nichols \cite{ nichols2019lessons}. \textbf{Unlike other pub-sub systems designed for TCP/IP or ICN, Sync-based pub-sub systems do not require central servers, changes in the network layer, or a name resolution system}.  
Instead, publishers and subscribers simply agree on a common name for the Sync group, and the subscribers will be notified by the Sync protocol of the new data names whenever a publisher publishes new data.

Given Sync's ability to support pub-sub models, we use it for distributed service discovery.  
Service publishers and finders can use a semantically meaningful name for their sync group.  e.g., \textit{``/edu/abc/library/printers/NDNSD/discovery"} for publishing and discovering the services of all the printers in a university library. Similarly, \textit{``/edu/abc/cs/ndnrepo/NDNSD/discovery"} can be used to discover NDN repositories in a computer science department. These semantic names from the application layer are directly used in the NDN network layer for rendezvous, thus there is no need for name resolution or central servers. 
Additionally, we use name-based access control (NAC-ABE)
to control the accessibility of sensitive services by an unauthorized user. The focus of our work is therefore threefold: (a) design the naming scheme and structure of service information (Section \ref{service-info}), (b) publish and discover the service information, and (c) Service-info accessibility. Figure \ref{fig:workflow} shows the workflow of NDNSD among several components of the system.
In the following sections, we detail the NDNSD protocol design.


\begin{figure}
    \centering
    \includegraphics[scale=0.31, trim= 0in 0.1in 0.2in -0.2in, angle=0]{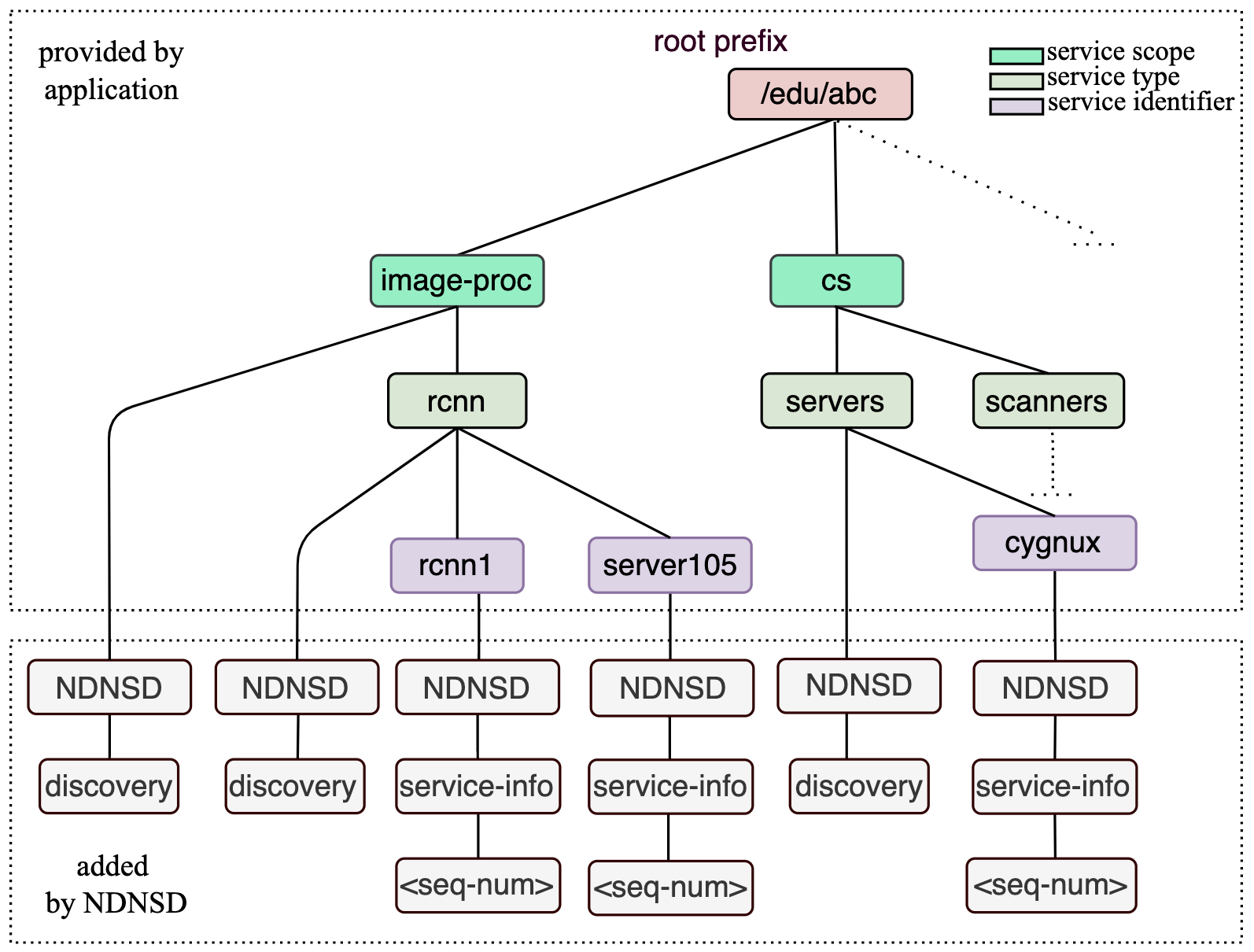}
    \caption{NDNSD Namespace Design}
    \label{fig: ndnsd-namespace}
    \vspace{-15 pt}
\end{figure}
\subsection{Hierarchical Namespace}
\label{Hierarchical-Namespace}




NDNSD uses two separate namespaces for discovery and application service information, and their design is illustrated in Figure~\ref{fig: ndnsd-namespace}

\subsubsection{Discovery Namespace}
Service publishers and finders use a Sync group to rendezvous with each other and we call the name prefix of their Sync group \emph{Discovery Name Prefix}. 
As shown in Figure~\ref{fig: ndnsd-namespace}, the discovery name prefix starts with a routable \textbf{root} component, which can be an organization, e.g., \textit{/edu/abc}, or an application, e.g., \textit{/app/mhealth}.
Depending on the specific application scenario, there can be more name components to further constrain the \textbf{service scope}, such as physical location (e.g., university library), organization entity (e.g., computer science department), and target users (e.g., computer science students).  The next component is \textbf{service type}.  For example, \textit{image-proc} represents all the image processing services, while \textit{image-proc/rcnn} includes only those running the RCNN algorithm. 
The above name components are provided by the application to NDNSD through its API (see Section~\ref{sec:API-protocol}). Finally, the last two name components are added by NDNSD to yield the discovery name prefix, e.g., \textit{/edu/abc/image-proc/rcnn/NDNSD/discovery}.  ``\textbf{NDNSD}'' differentiates NDNSD's data from other protocols' data, and ``\textbf{discovery}'' differentiates NDNSD's service discovery Sync data from its service-info data.

The above design does not impose a strict limit on the depth of the namespace hierarchy. Moreover, applications have the flexibility to use semantically meaningful names. Note, however, that service publishers and finders in one application need to agree on a service type and avoid collision with other applications. There may be well-known service types emerging as NDNSD is gradually adopted. 





\subsubsection{Service Information Namespace}
Each service provider publishes detailed information about its service in the relevant Sync group (see the previous section) so that service finders can choose the most suitable provider.  Figure~\ref{fig: ndnsd-namespace} illustrates the design of our service information namespace. It has the same first three name components, i.e.,  root, service scope (optional), and service type, as the discovery namespace. However, it requires a \textbf{service identifier} after service type, e.g., ``rcnn1'' after  \textit{/edu/abc/image-proc/rcnn}, to identify a specific service provider, which is supplied by the application.  The next two name components, ``NDNSD" and ``service-info", are added by NDNSD. 
For example, \textit{/edu/abc/cs/servers/cygnux/NDN/service-info} is the name prefix of the service information about the server \emph{cygnux} in the computer science department.  The last component is a sequence number for the service information -- whenever the server's information changes, the sequence number is increased.  The service provider will publish the new name through Sync, which will notify the service finders of the new name so they can fetch the new service information.



Note that service publishers should have a valid certificate for the name they want to use to advertise their service, as NDN requires every data packet to be signed by the public key of the data publisher.

\subsection{Service Information}
\label{service-info}

\begin{figure}
    \centering
    \includegraphics[scale=0.25, angle=0]{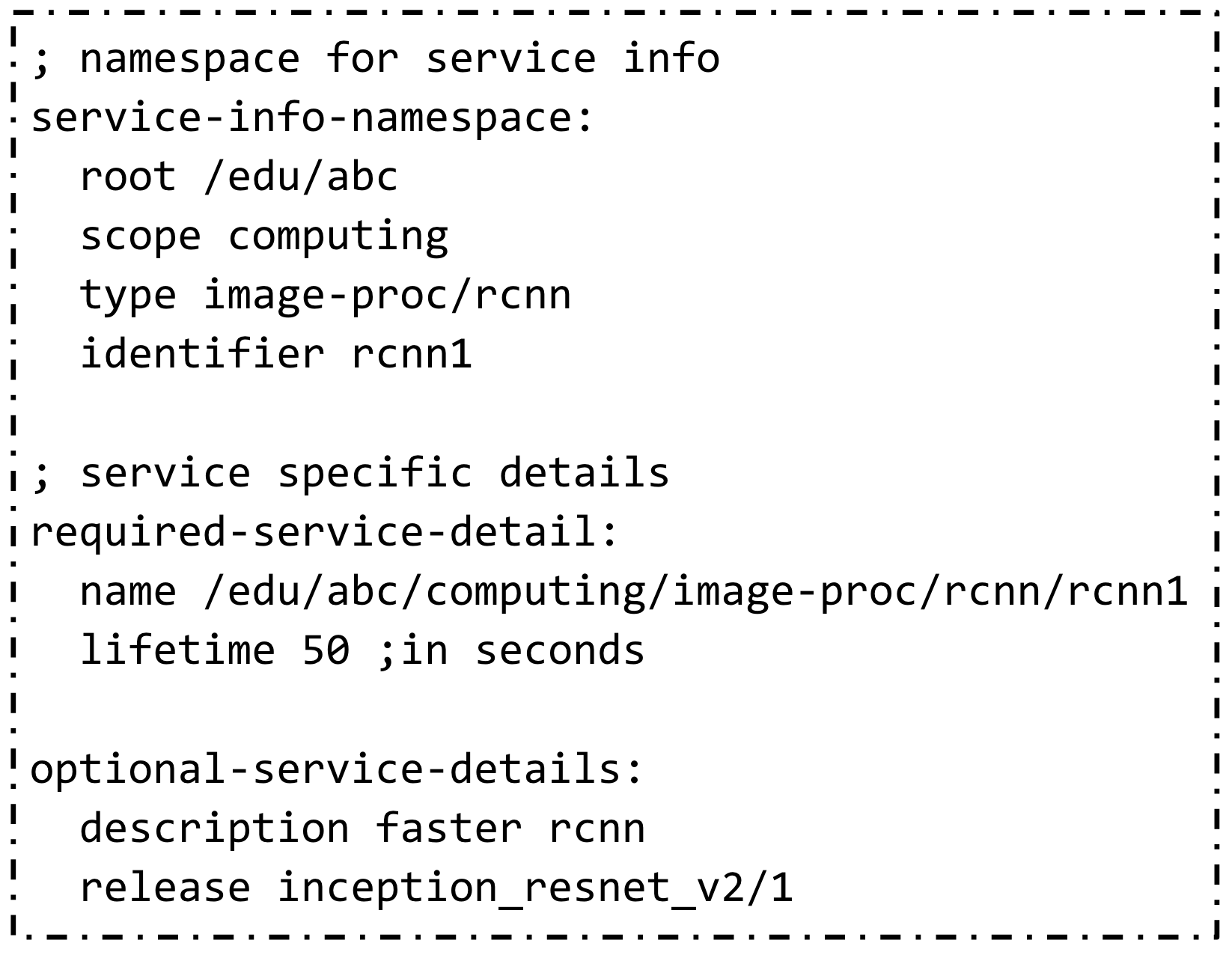}
    \caption[Sample Service Publisher Configuration File]{Sample Service Publisher Configuration File}
    \textsuperscript{(The above sample file is written in Boost Info Format \cite{kalicinski2008boost})}
    \label{fig: service-info}
    \vspace{-12 pt}
\end{figure}

Service information is a collection of information used by a service provider to advertise its service. 
Since NDNSD is a generic service discovery protocol, the service information design should support a wide variety of applications.
As shown in Figure~\ref{fig: service-info}, it is composed of three blocks: service info namespace, required service detail, and optional service detail. The items in the namespace block are used to construct discovery and service information name prefixes (Section~\ref{Hierarchical-Namespace}) to advertise the service.  
Using the two service detail blocks, providers can list details of their service using as many key-value pairs as needed. The required block contains the \emph{service name}, i.e., how to reach the service, and \emph{lifetime}, i.e., how long this service information is valid.  The optional block contains more application-specific information. For example, an image-processing service can list details about the model used to process the image and release version. IoT applications can provide details about sensor type, location, available memory, sleep time, processing capabilities, etc. 
\vspace{-10pt}
\subsection{API and Protocol Interactions}
\label{sec:API-protocol}

NDNSD provides the following API for applications to publish and discover service information:

\noindent \textbf{Service Publisher} receives and stores the service information from the application.  
It uses the root, service scope, and service type information in the namespace block to form a discovery prefix (e.g., \textit{/edu/abc/library/printers/NDNSD/discovery}) and joins the corresponding sync group. The publisher also uses the root, service scope, service type, and service identifier to construct the service-info data name, e.g., \textit{/edu/abc/printers/printer1/NDNSD/service-info/1}, creates a data packet with this name, and stores information from the service-detail blocks in the content of the packet.


\noindent \textbf{Service Finder} accepts service discovery requests from an application.  Each request contains root, service scope, and service type which are combined to form a discovery prefix. Next, the finder joins the sync group identified by the discovery prefix, fetches the service information from all the service providers, and sends it back to the application. In addition, the finder can provide measurement information
to the application on demand (Section~\ref{measurement}).
\vspace{-2pt}
\subsection{Measurement Information}
\label{measurement}

Measurement information is crucial for applications to determine the Quality of Service and perform load balancing.
In order to help service finders make better decisions in selecting service providers, NDNSD computes statistics such as round-trip time (RTT), retransmission count, and timeouts, while fetching the service information or probing a service provider using its service name. The measurements are exposed to the application via the Service Finder's API.

\subsection{Service-info Accessibility}
\label{service-info-accessibility}
\begin{figure}[h!]
    \centering
    \centering
    \includegraphics[width=0.5\textwidth]{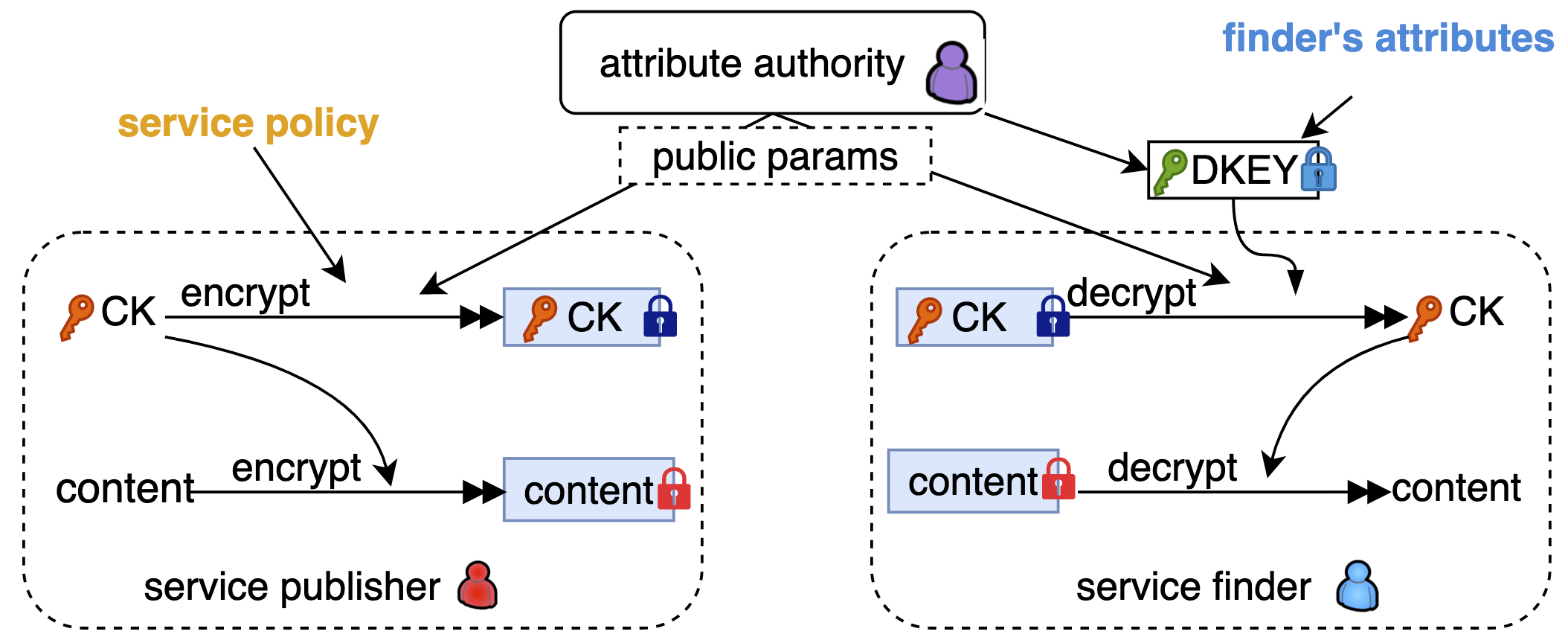}
    \caption[Use of NAC-ABE in Our System]{Use of NAC-ABE in Our System. Note: \textit{content} in the context of NDNSD is service information}
    \label{fig:name-based-access-control}
    \vspace{-10pt}
\end{figure}

Sensitive services may require some form of access control such that only authorized users can access their service information.  
For example, a printer in the CS department chair's office may want to restrict public use.
We use Name-Based Access Control with Attribute-based Encryption (NAC-ABE)~\cite{zhang2018nac} to control the visibility of service information. As shown in Figure~\ref{fig:name-based-access-control}, NAC encrypts the service information with policies composed of attributes, and generates decryption keys for only those with appropriate access rights. For example, the printer in the Chair's office can advertise a printer service and encrypt the service information with the policy (``CS Chair'' or ``CS Admin Assistant''). Now to decrypt this service information, the finder needs to have a decryption key with either the ``CS Chair'' or the ``CS Admin Assistant'' attribute.

Note that NAC is not used for all the advertised services by default. It is up to each service provider to decide whether to use NAC, based on the sensitivity of the services it offers. For example, public printers do not need access control on their service information.

%% file: evaluation.tex
\begin{figure*}%
\centering
\begin{subfigure}{.5\textwidth}
\includegraphics[width=\columnwidth]{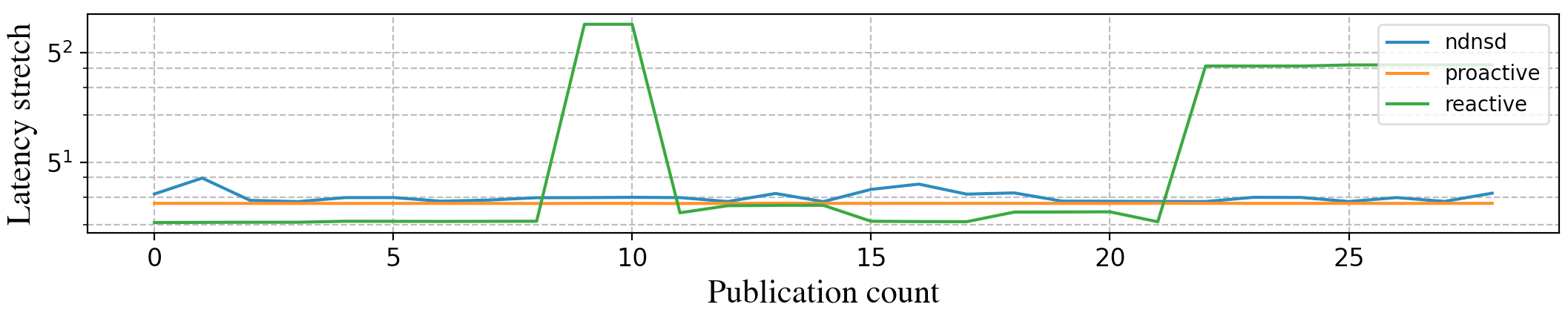}%
\caption{}%
\label{subfiga}%
\end{subfigure}\hfill%
\begin{subfigure}{.5\textwidth}
\includegraphics[width=\columnwidth]{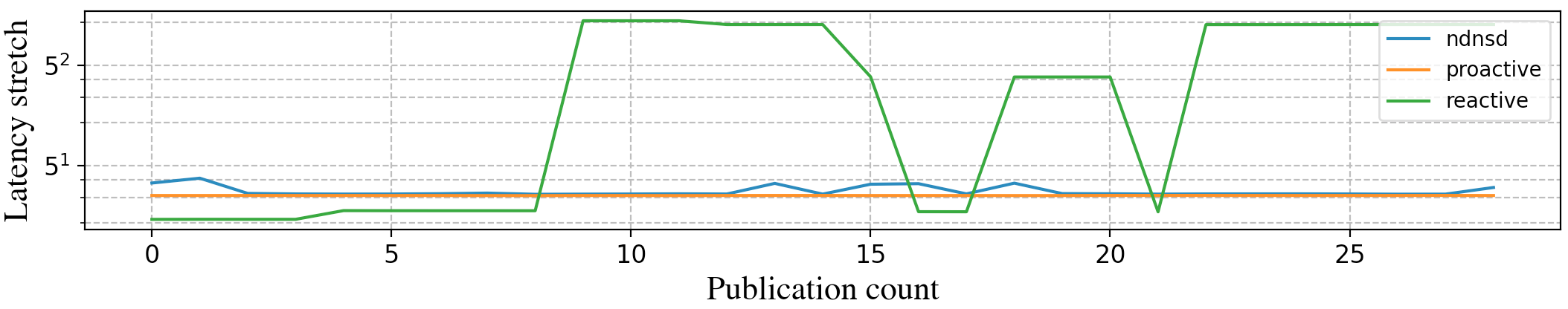}%
\caption{}%
\label{subfigb}%
\end{subfigure}\hfill%
\begin{subfigure}{.8\textwidth}
\includegraphics[width=\columnwidth]{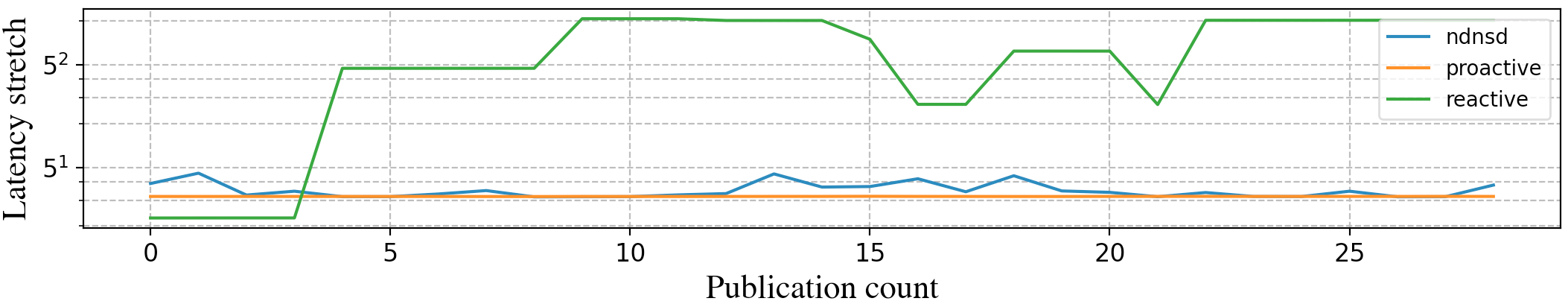}%
\caption{}%
\label{subfigc}%
\end{subfigure}%
\caption{Figure (a), (b), and (c) shows the median, 75th, and 90th-percentile service discovery latency stretch, respectively, for each successive publication}
\label{fig:stretches_graph}
\vspace{-12 pt}
\end{figure*}

\section{EVALUATION}
\label{evaluation}

We evaluate NDNSD by comparing it with two simple service discovery schemes, one is \textbf{proactive} and the other is \textbf{reactive}. In the \emph{Proactive} scheme, the service provider periodically multicasts its service to the entire network using a notification Interest. 
The Interest consists of the discovery prefix (e.g. \textit{/edu/abc/cs/servers}), the application service name (e.g. \textit{/edu/abc/cs/servers/cygnux/service-info}), and a sequence number. The sequence number is increased whenever the service status is updated. 
An application interested in the service listens to the multicast, uses the application service name to construct a service-info Interest, and fetches the corresponding service-info using unicast. The sequence number tells the finder if the multicast was already received and, if so, the finder avoids fetching the same service-info. In this scheme, multicast Interests from multiple service providers (of the same service) are not aggregated due to the presence of the application service name. This can significantly increase packet overhead with an increasing number of providers.
In the \emph{Reactive} scheme, the finder multicasts service discovery Interests to the network. Since there can be more than one service provider, the finder iteratively sends multicast Interests, with  known providers carried in the application parameter field.  This process continues until the Interest times out, which means the service-info from all the providers have been fetched. Each provider, after receiving the multicast Interest, checks if its name is present in the application parameter. If so, it ignores the Interest. Otherwise, it sends back its service-info. This iterative process is also periodic, which helps the finder fetch all the updates from the providers. 
For a fair comparison, we set the frequency of periodic Interest multicasts for all the schemes to once every second in our experiments. \\
\noindent \textbf{Setup} For our evaluation, we emulate multiple service providers offering \textit{Computation Service}. Each provider updates its service-info, such as available GPUs, TPUs, Disk Storage, and Server-load, every 10 seconds for a total of 30 updates. 
We use Mini-NDN~\cite{mini-ndn}, a lightweight network emulator tool, and the NDN testbed topology\footnote{NDN testbed topology: http://ndndemo.arl.wustl.edu/} consisting of 37 nodes and 97 links 
for all experiments. 

\subsection{Performance Metrics}
\vspace{-3pt}
\noindent We use the following metrics in our evaluation:

\textbf{a) Service Discovery Latency Stretch} is the ratio between the actual service-info latency (i.e., time to discover some service-info data) and the expected minimum of the same. 


\textbf{b) Normalized Packet Overhead} (per node) is the ratio between the overhead per node and the expected minimum number of packets per node. For example, suppose there are 5 service finders and 2 service providers each publishing 30 times at an interval of 10 seconds, 
the expected minimum number of packets = number of providers $\times$ number of links in both directions $\times$ number of publications =  2 $\times$194 $\times$ 30 = 11640. For NDNSD, the overhead consists of (i) Sync Interest and Data packets used by the service providers and finders for synchronization, and (ii) the duplicate NACKs sent by the nodes to notify the downstream of receiving the same Interest twice. For the Proactive scheme, the overhead consists of periodic service multicast Interests from the service providers and the duplicate NACKs. For the Reactive scheme, the overhead consists of periodic multicasts to find service from the finders and the duplicate NACKs. Additionally, for all the schemes, extra service-info Interests and data packets are also counted as an overhead. Ideally, there should only be one Interest and one Data packet for each publication. 

\textbf{c) Satisfaction Ratio} for a specific service is defined as the number of service providers learned by a finder divided by the total number of service providers offering the same service.

\subsection{Emulation Results}
In Figure \ref{fig:stretches_graph}, we present the median, 75th, and 90th-percentile of the \textit{Service Discovery Latency Stretch}. The results show that both NDNSD and the Proactive scheme have similar low stretches.
This is because in both schemes, the service provider advertises its service to the entire network after it is published or updated. The advertisement helps the finders to fetch the service-info immediately after receiving the multicast. Figure \ref{fig:stretches_graph} also shows that the Reactive scheme performed worst among the three, because in this scheme the publisher needs to wait for a multicast Interest from the finder to send back the service-info. Thus, delayed arrival of the Interests will increase the service-info latency, which explains its high stretch. Occasionally, in the Reactive scheme, Interest arrival and service advertisement can happen at the same time, resulting in a short service-info latency equal to the delay between the service provider and the finder. This can be seen in Figure \ref{fig:stretches_graph}(a) \& (b) where the Reactive stretch is lower than the other two schemes and close to 1.
 \begin{figure}
    \centering
    \includegraphics[scale=0.27, angle=0]{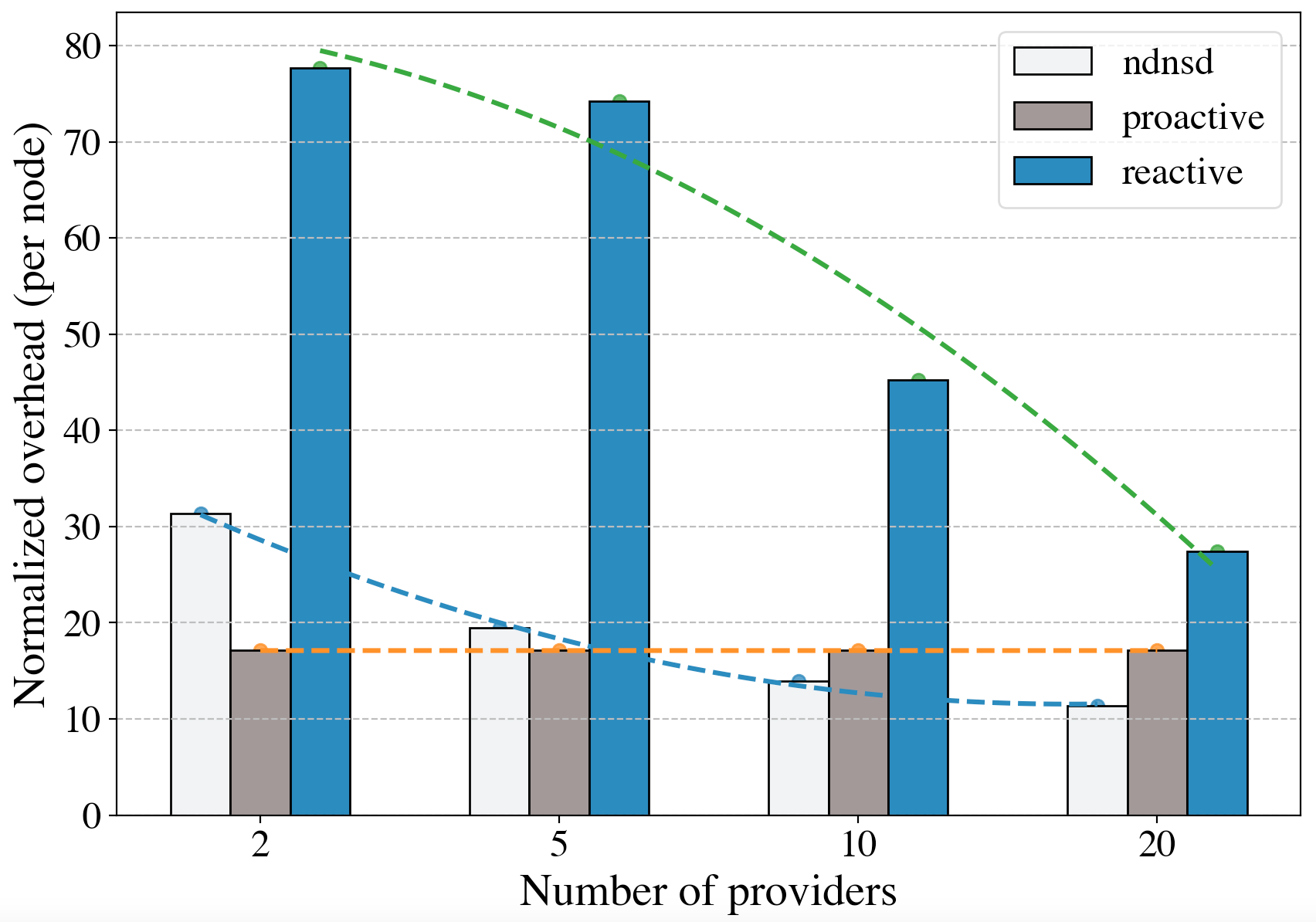}
    \caption[Normalized Overhead]{Normalized Packet Overhead}
    \label{fig: overhead-norm}
    \vspace{-18 pt}
\end{figure}

In Figure~\ref{fig: overhead-norm}, we present the \textit{Normalized Packet Overhead} of the different schemes. The overhead in the Proactive scheme remains constant as the number of providers increases because the expected and actual number of packets increased by the same factor due to the lack of Interest aggregation in the Proactive scheme. It can be seen that for the other two schemes, NDNSD and Reactive, the overhead starts decreasing as the number of providers increases. This is because of the Interest aggregation in both schemes. 
In addition, NDNSD performed much better than the Reactive scheme.
This is because: (a) NDNSD fetches service-info using unicast whereas the Reactive scheme uses multicast, (b) The Reactive scheme always uses one extra final multicast Interest, which times out, to make sure service-info from all the providers are fetched, and
(c) a greater number of multicast packets in the Reactive scheme leads to more duplicate NACKs. Thus, we can expect that NDNSD will perform better than the Reactive scheme, even if the number of publishers grows much higher. 
 
Finally, we ran a set of experiments introducing 1, 2, and 4 percent link loss per link and compared the \textit{Satisfaction Ratios}. We observed that both NDNSD and the Reactive scheme were able to receive all the service-info data regardless of the link losses, thereby maintaining a 100\% satisfaction ratio. For NDNSD, Sync actively synchronizes the publications using the states among all participating nodes. 
Thus, advertisements from all the service providers reach the finder. Once an advertisement is received, NDNSD fetches the service-info and does multiple retransmissions if needed.  In the Reactive case, the finder's periodic discovery Interest is able to retrieve/recover all service-info data.  In contrast, in the Proactive case, there is no finder retransmission, so lost service-info packets cannot be recovered. Hence, we observed 99\%, 97.5\%, and 97\% satisfaction ratios for 1, 2, and 4 percent link loss, respectively.

%% file: conclusion.tex
\section{Conclusion and Future Works}
\label{conclusion-future-work} 
\vspace{-1pt}
We have presented NDNSD, a fully distributed, general-purpose protocol for service publishing and discovery in NDN.  We use a hierarchical namespace for NDNSD, which provides applications with fine-grained control over the advertised names of their service(s).
Moreover, we leverage NDN Sync to make service discovery independent of any external infrastructure, and designed the service information to support a wide variety of applications.
NDNSD provides measurement information to applications to allow better decision-making when selecting service providers, and is able to control the accessibility of service information using NAC.  Finally, we have shown that NDNSD outperforms two other baseline service discovery solutions.


We plan to do the following in our next steps: 
(a) evaluate NDNSD's performance in applications such as building management systems and improve our design and implementation accordingly,
(b) refine the collection and representation of measurement information,
and (c) conduct evaluation over NDN testbed and release NDNSD to the public.
\vspace{-3pt}

\section*{ACKNOWLEDGMENT}
This work was supported by the National Science Foundation awards 1629769 and 2019085. We thank the anonymous reviewers for their insightful feedback.